\documentclass[referee]{cjaa}           

\usepackage{graphicx}                   
\input{epsf.sty}                        
\input{psfig.sty}                       

\begin{document}

\title{Spectral properties of anomalous X-ray pulsars
}


   \author{Y. Lu \mailto{}, W. Wang and Y.H. Zhao
      }
   \offprints{Y. Lu}                   

   \institute{National Astronomical Observatories, Chinese Academy of Sciences,
             Beijing 100012, China\\
             \email{ly@bao.ac.cn}
          }



   \date{Received~~2003 month day; accepted~~2003~~month day}

   \abstract{
In this paper, the spectra of the persistent emission from
anomalous X-ray pulsars (AXPs) and their variation with spin-down
rate $\dot{\Omega}$ is considered. Firstly, based on an
accretion-powered model, the influences of both magnetic field and
mass accretion rate on the spectra properties of AXPs are
addressed. Subsequently, the relation between the spectral
property of AXPs and mass accretion rate $\dot{M}$ is investigated.
The result shows that there exists a linear correlation between
the photon index and mass accretion rate, and the spectral
hardness increases with increasing $\dot{M}$. A possible emission
mechanism for the explanation of spectral properties of AXPs is
also discussed.
}

   \authorrunning{Y. Lu }            
   \titlerunning{anomalous X-ray pulsars}  

   \maketitle
%
%
\section{Introduction}           
\label{sect:intro}

Anomalous X-ray pulsars (AXPs) are one of the enigmatic class of
Galactic high energy sources. These sources differ from known
magnetic accreting X-ray pulsars found in high and low mass X-ray
binaries (HMXBs and LMXBs) (Mereghetti \& Stella \,1995; Van
Paradijs et al. \,1995). AXPs are sources of pulsed X-ray emission
with spin periods in the 6-12s range, very soft X-ray spectra,
secular spin down on time scales of $\sim 10^3-10^5\,yr$, and lack
of bright optical counterparts. Two or possibly three of AXPs are
close associated with supernova remnants (SNRs). Additionally,
AXPs share some similarities with the Soft Gamma-ray Repeaters
(SGRs) (Hurley\,2000).

Two broad classes of models have been proposed to understand the
X-ray emission from AXPs. The first class of models assume that
the sources of AXPs are isolated neutron stars with ultra-magnetic
field strengths in the range $10^{14}-10^{15} \rm {Gauss}$ - i.e.
``magnetars'' (Thompson \& Duncan \,1996; Hely \& Hernquist
\,1997). The spin-down of the pulsars is primarily due to magnetic
dipole radiation, then the AXPs have enormous surface magnetic
dipolar fields. The second class of models suggested to explain
AXP emissions is that they are accreting from a disk of material
leftover from the supernova explosion that created the neutron
star (Chatterjee et al.\,2000, hereafter CHN; Alpar\,2001; Lu \&
Cheng \,2002; and Marsden et al.\,2001). Such models do not
require neutron stars with unusually strong magnetic fields. In
these cases, the spin-down torque is external and, supposedly, a
neutral consequence of accretion braking (Francischell \& Wijers
\,2003). A very different scenario, based on strange
matter stars, has been proposed by Dar \& DeRujula \,(2000). They
argued that AXPs are either strange stars or quark stars in which
the X-rays are powered by gravitational contraction, and the
spin-down is due to the emission of relativistic jets.

AXPs could be magnetars, but valid alternative models cannot be
ruled out with a decisive proof (Mereghetti et al.\,2002). In the
magnetar model, a more stable luminosity, spectrum and $\dot{P}$
noise can be expected more naturally than in accretion-powered pulsars. The
spectra and light curves expected from the surface of highly
magnetized neutron stars have been computed by several authors
(\"Ozel \,2001; Zane et al. 2001; Ho \& Lai 2001). The spin-down
of magnetars is due to a combination of standard magnetic dipole
radiation torque and torque from the wind (Harding et al. 1999).
Both of these torques increase strongly with magnetic field
strength, and therefore the hardening of the power-law spectral
component with $\dot\Omega$ implies a similar hardening of the
underlying Alfven wave spectrum with increasing magnetic field
$\rm {\bf B}$. \textit{It is known that the observation of AXPs
shows that the spectral hardness decreases with increasing
spin-down rate (Marsden \& White \,2001). Therefore, the
correlation of the spectral hardness (photon index $\Gamma$) and
the spin-down rate ($\dot{\Omega}$) predicted by magnetar model is
inconsistent with the data.}

Based on the above investigations, alternatively, we relate
accreting neutron stars depending on both the magnetic field and
the accretion rate $\dot{M}$ to model the spectral properties and
spin-down rates of AXPs. The purpose of the present paper is
twofold. Firstly, based on the observation, the relation between
the photon index $\Gamma$ of AXPs/SGRs and the accretion rate
$\dot{M}$ is investigated. Secondly, a possible emission mechanism
is suggested to model the spectral properties of AXPs/SGRs.
Finally, some predictions based on the relation between
$\Gamma-\dot{M}$ are discussed.

\section{Parameters and observational data for AXPs}
\label{sect:Obs}

An important parameters in describing
AXPs is its spin period ($\rm{P}$). It is known that the AXPs slowly brake,
so the time derivative of the period, $\dot{P}$, is another one. In addition,
based on two observations, other useful pulsar parameters
can be defined: the ``characteristic age" $\tau=P/2\dot{P}$, and
the magnetic field
\begin{eqnarray}
 B=\left(\frac{3Ic^3}{8\pi^2r^6}\right)^{1/2}(P\dot{P})^{1/2}\,\,\,,
 \label{eq:B-dotp}
\end{eqnarray}
where $I\sim 10^{45}\rm {g}\,\rm{cm^2}$, and $r\sim 10^6
\rm{cm}$.

The role of period derivatives $\dot{P}$ is confirmed when
observing the AXPs spectral index variation. Following
the data analysis of Marsden \& White (\,2001) for the
phase-averaged spectra of the AXPs and SGRs in the range of
$0.5-10.0$ keV, we obtained the basic data of all SGRs and AXPs with known spin periods and period
derivatives detected with ASCA which is summarized in Table\,\ref{Tab:publ-AXP}.
The period derivative of one object
AXP 1048-59 cannot be determined definitely, so we take two values
throughout the following analysis. And the photon indexes here are
derived in terms of two-component black-body plus power law
spectral models. Figure\,\ref{Fig:plot1} shows the variation of
the photon index $\Gamma$ vs. spin-down rate $\dot{\Omega}$ for
such AXPs and SGRs.

The relation of the photon index
$\Gamma$ and $P, \dot{P}$ is fitted in terms of the linear equation
$\Gamma=\alpha \log P+ \beta \log\dot{P}+\gamma$. For five AXPs and two SGRs listed in Table\,\ref{Tab:publ-AXP},
the fitting of $\Gamma(P,\dot{P})$ is
\begin{eqnarray}
\Gamma(P,\dot{P})=2\log P-\log\dot{P}_{12}+2.5\,\,\,,
\label{eq:Gamma-dotp}
\end{eqnarray}
 where $\dot{P}_{12}$ denotes $\dot{P}/10^{-12}{\rm s s^{-1}}$.
 Substituting $\Omega=2\pi/P$ and $\dot\Omega=2\pi\dot{P}/P^2$ into
Eq.~\ref{eq:Gamma-dotp}, the correlation between $\Gamma$ and
$\dot\Omega$ can be obtained. $\Gamma$ is a linear function of
$\dot\Omega$. The photon index $\Gamma$ decreases with increasing
spin-down-rate $\dot{\Omega}$. (see Fig.~\ref{Fig:plot1}).

Furthermore, the relation (\textit{($\Gamma-B$)) between the photon index and
magnetic field strengths for such AXPs/SGRs is studied. We plot it in Figure
\,\ref{Fig:plot2}. Figure 2 shows that the harder the spectra of
AXPs/SGRs, the stronger the magnetic fields}. It seems that the
present magnetar model cannot give this result yet, so it may
require the further theoretical considerations. The following
study shows an accretion-based model (CHN) could give a possible
explanation to the observed correlation between the photon index
and spin-down rate in AXPs and SGRs.
\section{The spectral properties of AXPs and a possible model}
\subsection{The diagram of spectral photon index and the accretion rate ($\Gamma-\dot{M}$) }
 \label{subsect: models}

In accretion-based models, a neutron star can be observed as an
X-ray pulsar during a tracking phase. AXPs are supposed to rotate
at a quasi-equilibrium period. And a spin-down torque $\dot{J}$ is
needed during this period. CHN suggests $\dot{J}$ changes with time according to
the following formula\,(Menou et al. \,1999)
\begin{eqnarray}
\dot{J}=I\dot{\Omega}=2\dot{m}R^2_m\Omega_K(R_m)\left[1-\frac{\Omega(t)}{\Omega_K(R_m)}\right]\,\,\,,
\label{eq:dotJ-dotOmega}
\end{eqnarray}
where $\Omega_K(R_m)$ is the Keplerian rotation rate at radius
$R_m$, and $R_m$ is the magnetospheric radius which is determined
by
\begin{eqnarray}
 R_m\approx 6.6\times
10^7B_{12}^{4/7}\dot{m}^{-2/7}\,\,\,\rm{cm}
 \label{eq:Rm-dotm}
\end{eqnarray}
In Eq.\,\ref{eq:Rm-dotm}, $\dot{m}=\dot{M}/\dot{M}_{Edd}$, where
$\dot{M}_{Edd}\approx 9.46\times 10^{17} \,{\rm g}\,{\rm s^{-1}}$
is the Eddington accretion rate, $B_{12}=B/10^{12}\,{\rm Gauss}$.

Following the arguments of Cannizzo et al.\,(1990), CHN suggests
that after a dynamical time $T$ the fossil disk loses mass
self-similarly,
\begin{eqnarray}
\dot{m} =\left\{ \begin{array} {r@{\quad;\quad}l}
\dot{m}_0 \,,\,\,\,\,\,\,\,\,\,\,\,\,0<t<T\,\,\,\\
\dot{m}_0\left(\frac{t}{T}\right)^{-\alpha}\,,\,\, t\ge T
\end{array}
\right.
\label{eq:dotm-t}
\end{eqnarray}
where $T\sim 10^{-3}\,{\rm s}$, is the local dynamical time, and
$\dot{m}_0$ is a constant, which is normalized to the total
initial disk mass, $M_d=\int^8_0\dot{M}_d dt$, by
$\dot{m}_0=[(\alpha -1 )M_d]/(\alpha\dot{M}_{Edd}T)$. $\alpha>1$
is a constant that depends on directly the disk opacity
(Francischell \& Wijers \,2003). Assuming an arbitrary value for
$\alpha$, equation \,\ref{eq:dotm-t} and \ref{eq:dotJ-dotOmega}
can be combined to yield an analytic formula for ${\Omega(t)}$, in
terms of incomplete gamma functions (CHN). For $\alpha=7/6$, the
solution of equation ~\ref{eq:dotJ-dotOmega} is
\begin{eqnarray}
\Omega(t) =\left\{ \begin{array} {r@{\quad;\quad}l}
\Omega_k(R_m,0)\dot{m}_0^{3/7} \,,\,\,\,\,\,\, 0<t<T\,\,\,\\
\Omega_k(R_m,0)\left(\frac{\dot{m}}{\dot{m}_0}\right)^{3/7}\,,\,\,\,\,
t\ge T
\end{array}
\right.
\label{eq:dotOmega-t}
\end{eqnarray}
Since $\Omega_k(R_m,0)=GM_{ns}/R_m^3$, then for $t\ge T$
\begin{eqnarray}
\Omega(t)&\approx& C_1B_{12}^{-6/7}\dot{m}^{3/7}\,\,\,,
\label{eq:dotOmega2-t}
\end{eqnarray}
where $C_1$ is a constant, and $C_1=25.48$. If $\Omega(t)$ evolves
as equation\,\ref{eq:dotOmega2-t}, the spin period of the star
nearly equals the evolving equilibrium period; this is a tracking
phase (CHN). It is worth noting that the spin periods of SGRs
and AXPs implied by equation\,\ref{eq:dotOmega2-t} depend on both
magnetic fields and mass accretion accretion rates.

A strong propeller torque is needed to determine the spin
evolution of the neutron star + fossil disk system (CHN). However,
the way in which a propelling neutron star loses angular momentum
is not well understood (Francischell \& Wijers \,2003). In the
most general form, the exact manner in which angular momentum is
transferred between an accretion disk and a neutron star is a
complex magneto-hydrodynamical problem, so there is no simple
analytic solution for $\dot{\Omega}$ changing with $\dot{m}$ and
$B$ has been determined from equation 7. It is thought that most
of the captured matter is ejected prior to reaching the neutron
stars surface, i.e., in the strong propeller phase. However, after
a sharp propeller cycle, the system reaches a tracking phase which
is argued as producing ``AXPs" phase (CHN). Clearly, employing
different propeller torque  (Francischell \& Wijers \,2003; Fabian
1975; and CHN) will produce different results in determining the overall
evolution of the system. How the propeller torques affect the
period changing with time has been illustrated by Francischell \&
Wijers (2003). \textit{Based on above discussions, it is safe to
assume that some kind of propeller torque acts on the neutron star
between the strong propeller phase and tracking phase in order to
spin it down with time. So, for simplicity, in this paper, we
assume that the propeller torque is given by},
\begin{eqnarray}
\dot{J}=I\dot{\Omega}=-2\xi\dot{m}R^2_m\Omega(t)\,\,\,.
\label{eq:dotJ-dotm}
\end{eqnarray}
\textit{where $\xi$ is a constant, which is in the range of
 $0<\xi\leq 1$. $\xi=1$ is for the strong propeller limit
(Francischell \& Wijers \,2003), and $\xi\sim 0$ is for the
quasi-equilibrium phase, i.e., tracking phase. Then, from}
equation\,\ref{eq:dotJ-dotm}, we obtain
\begin{eqnarray}
\dot{\Omega}&\approx&C_2B_{12}^{2/7}\dot{m}^{6/7}\,\,\,,
\label{eq:dotOm-dotm}
\end{eqnarray}
\textit{where $C_2=2.164\times 10^{-10}\xi$ is a constant}.

Combining equation\,\ref{eq:dotOmega2-t} and \ref{eq:dotOm-dotm}, and
taking into account of $\Omega=2\pi/P$ and $\dot\Omega=2\pi\dot{P}/P^2$,
the mass accretion rate as function of $\dot{P}$ and $P$ is yielded
\begin{eqnarray}
\log \dot{m}(P,\dot{P})=-\frac{7}{3}\log P
+\log\dot{P}+[10.26-\log \xi ]\,\,\,. \label{eq:dotm-dotp}
\end{eqnarray}
Comparing equations \,\ref{eq:dotm-dotp} with
\ref{eq:Gamma-dotp}, a inverse relation between the photon index
$\Gamma$ and the mass accretion rate
$\dot{m}$ is inferred. Fitting the mass accretion rates $\dot{m}$ of 5 AXPs and 2 SGRs
estimated according to Eq.\,\ref{eq:dotm-dotp} in the cases of \textit{$\xi=1, \, 0.5$, and $0.05$,
respectively}, we obtain
\begin{eqnarray}
\log \dot{m}=A-0.99\Gamma\,\,\,. \label{eq:dotm-dotGamma}
\end{eqnarray}
\textit{where $A$ is a fitting constant, and $A=0.23, 0.53$ and
$1.53$ corresponding to $\xi=1, 0.5$ and $0.05$, respectively.
Equation 11 shows a good correlation between $\Gamma$ and
$\dot{m}$. The diagram of $\Gamma$-$\dot{m}$ in the case of $\xi=0.5$} is plotted in
Fig.~\ref{Fig:plot3}.

\subsection{A Possible mechanism for the emission of AXPs}
Figure ~\ref{Fig:plot3} implies that larger accretion rate
produces the harder spectra (high-energy tail) in accretion-based
model. The origin of this high-energy tail is unexplained very
well at present. However, this high-energy tail could be due to
thermal Comptonization by a hot coronal plasma, or it could be due
to non-thermal emission (Tavani \& Liang \,1996). Based on the
energetics of particle acceleration and cooling near the Alfven
radius, B\"ottcher \& Liang \,(2001) simulated the
thermal-nonthermal radiation from a neutron star with both a
weakly magnetized ($\sim10^9$G) magnetospheric accretion shell and
a normal magnetic case ($\sim10^{12}$G). The basic conclusion of
B\"ottcher \& Liang \,(2001) is: there is a positive correlation
between spectral index and accretion rate, i.e., the spectral
hardness and accretion rate are anti-correlated. It is stressed
that this conclusion is obtained in the situation of a neutron
star with a weakly magnetic field, i.e., $\sim10^9\,G$. The
interesting calculation for the photon spectra with the mass
accretion rate is for the normal magnetic case ($\sim10^{12}$G),
which are plotted in the figure 3 and figure 5 of B\"ottcher \&
Liang \,(2001). These two figures show that harder photon spectra are
produced for higher accretion rates, particularly around energy
10\,\rm{keV}\,(2-10\,\rm{keV}, ACSA energy bands) in which we are
concerned in this paper. As a natural consequence, in the case of
a neutron star with a normal magnetic field, the photon index
decreases with increasing accretion rates. In other word, there is
a negative correlation between spectral index and accretion rates
as observed in AXPs/SGRs.  Furthermore, the hard X-ray spectral
index resulted from the simulation of B\"ottcher \& Liang (2001)
are in excellent agreement with the values of $\Gamma\sim 3-4$
generally observed in AXPs. From this context, the fitting of
$\Gamma-\dot{m}$ shows that the accretion rate is in the range from
$10^{-2}$ to $10^{-4}$, which is consistent with the model
conditions suggested by B\"ottcher \& Liang \,(2001). Therefore,
we can apply the emission mechanism of B\"ottcher \& Liang
\,(2001) for the normal magnetic case to construct the
accretion-powered emission model for AXPs,  which can explain the
relation between the accretion rate and spectral properties.

\section{Discussions and Conclusions}
\label{sect:analysis}

\begin{figure}
   \begin{center}
    \psfig{figure=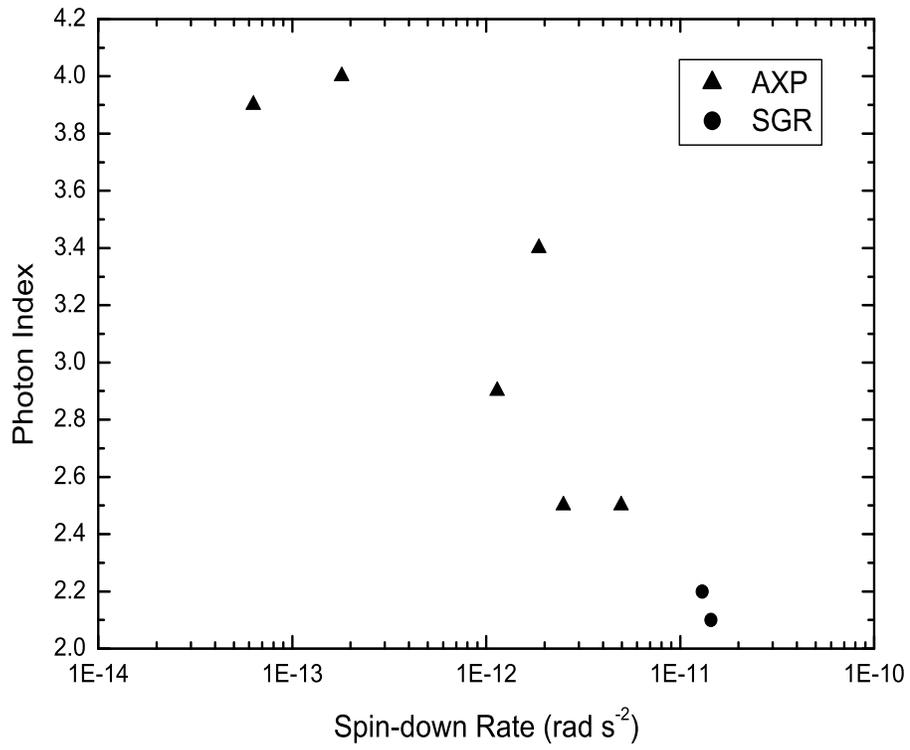,width=120mm,height=100mm,angle=0.0}
     \caption{The $\Gamma-\dot{\Omega}$ diagram of AXPs and SGRs.
 The triangles denote the AXPs and the circles are SGRs, and the symbols are all the same in the following figures.}
   \label{Fig:plot1}
   \end{center}
\end{figure}

\begin{figure}
   \begin{center}
 \hspace{3mm}
 \psfig{figure=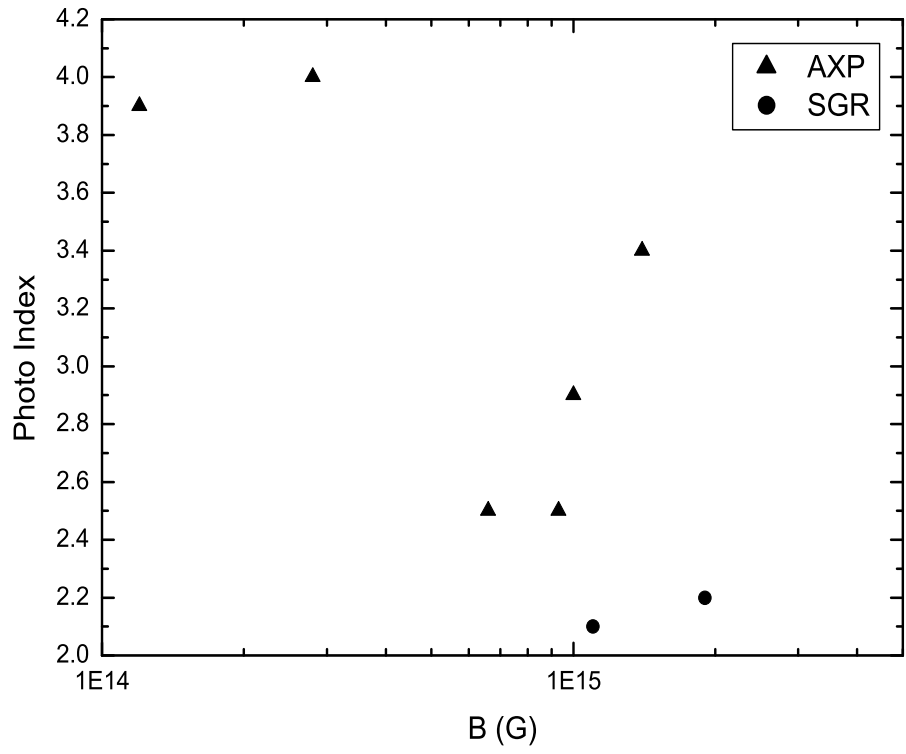,width=120mm,height=100mm,angle=0.0}
   \parbox{180mm}{{\vspace{2mm} }}
 \caption{The $\Gamma-B$ diagram of AXPs and SGRs.}
   \label{Fig:plot2}
   \end{center}
\end{figure}

\begin{figure}
   \begin{center}
 \hspace{3mm}
  \centering
 \psfig{figure=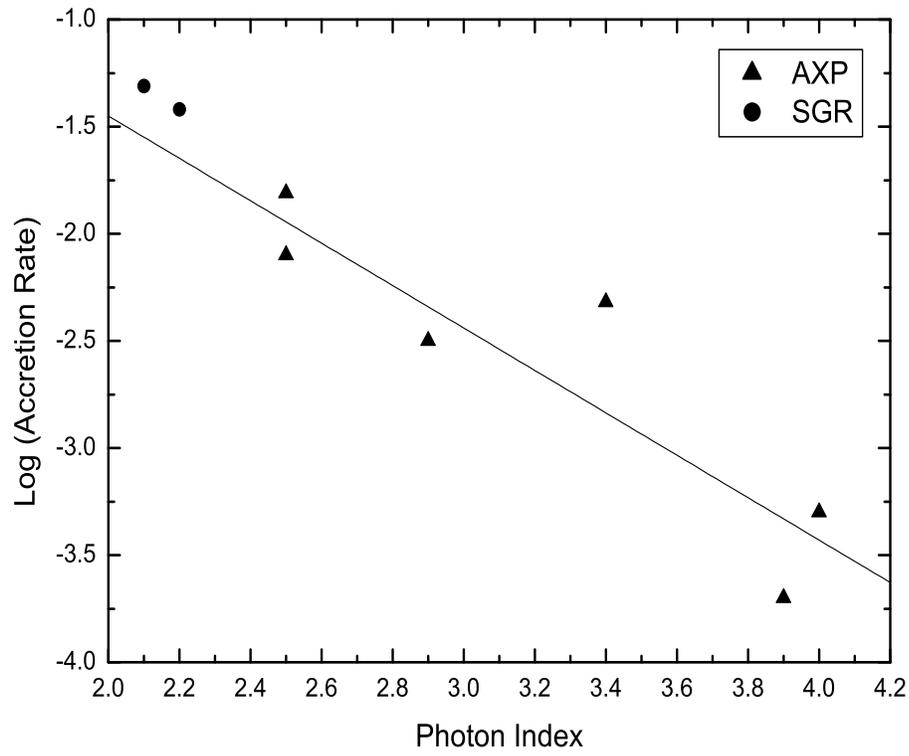,width=120mm,height=100mm,angle=0.0}
   \caption{The $\Gamma-\dot{M}$ diagram of AXPs and SGRs in the accretion model. The line denotes
  the best fiiting relation of all the data points.}
   \label{Fig:plot3}
   \end{center}
\end{figure}

The correlation between the spectra and the spin-down rate of
AXPs/SGRs shows that the photon index decreases with increasing
spin-down rate for AXPs and SGRs (see Fig.~\ref{Fig:plot1}).
This fact has not so far been given a good explanation. If the
increasing power-law emission with spin-down rate is consistent
with the magnetar model, the spectra of AXPs/SGRs is expected to
extend into the far-UV band (Marsden \& White(\,2001). This
implies that observation of spectral breaks in the non-thermal
persistent emission in the far-UV would be important evidence in
support of the magnetar model. With respect to the accretion model
(CHN), the spectral photon index and the accretion
rate $\dot{m}$ (see Fig.~\ref{Fig:plot3}) is investigated. \textit{To simplify the
calculating of our model, an alternative propeller
torque (see, Eq.(8)) to limit the spin-down rate at the boundary
when the system transfers from the propeller phase to ``tracking"
phase is introduced.} The analysis demonstrates that the correct spectral shape
for the values of $B$ and $\dot{m}$ can produce the rapid
spin-down of AXPs and SGRs. \textit{A relation of $\log
\dot{m}=0.53-0.99\Gamma$ is derived for the case of $\xi=0.5$.
This relation plotted in figure 3 shows that the hardness of the spectra increases with
increasing mass accretion rate. The same conclusion remains for
different values of $\xi$, only the slope of fitting lines is changed
because $\xi$ modifies the values of $\dot{m}$ (see Eq.(11)). In
principle, the} result shown in this paper is consistent with the model prediction
given by B\"ottcher \& Liang (2001) in the normal magnetic
situation. Consequently, it is possible that even stronger magnetic-field
accreting neutron stars with low-mass accretion rate may be
consistent with the data of AXPs/SGRs.

 Unlike the case for normal accretion onto neutron stars , the
observed relationship between the absolute X-ray luminosity and
the photon index $\Gamma$ in AXPs/SGRs cannot be inferred from the
relation of $\dot{m}-\Gamma$. This is because the X-ray luminosity
of the system is determined by the mass accretion rate on to the
surface of the star, which could be different from the mass
accretion rate through the disk (CHN). Two reasons can be involved
to account for this. Firstly, in the propeller phase, it is not
clear at all how much material eventually reaches the neutron star
surface, and what is the radiation efficiency for the matter
stopped  the neutron star surface. Secondly, the X-ray luminosity
may not cover the total luminosity.

\begin{acknowledgements}

We are particularly grateful to the anonymous referee for
insightful comments and suggestions. We would like to thank Ling
J.R. and Zhang W.M. of Tsinghua University for useful discussions
for useful discussions. This work was
supported by the National Natural Science Foundation of China
under grant no.10273011, the National 973 Project (NKBRSF
G19990754), and the Special Funds for Major State Basic Research
Projects.
\end{acknowledgements}

\clearpage
\begin{table*}[]
  \caption[]{ SGR AND AXP TIMING PARAMETERS
   }
  \label{Tab:publ-AXP}
  \begin{center}
  \begin{tabular}{@{}llllll@{}}
  \hline
   Object & Start date & $P(s)^a$ & $\dot{P}(10^{-12}{\rm s s^{-1}})^b$
    & $\Gamma\,(BB+PL)^c$ & ${\rm References}^d$\\
  \hline
 SGR 1900+14...... & 1998 Apr 30 &5.158971(7) &61.0$\pm$ 1.5& 2.1 & 1, 2  \\
                   & 1998 Sep 16 &5.16025(2)  &61.0$\pm$ 1.5 & \  & 2, 3\\
 SGR 1806-20...... & 1993 Oct 10 &7.468514(3) &115.7$\pm$ 0.2 & 2.2 & 4, 5  \\
                   & 1995 Oct 16 &7.46445(3) &115.7$\pm 0.2$& \  &4, 5\\
 AXP 1048-59...... & 1994 Mar 03 & 6.446646(1) & 32.9$\pm$0.3 & 2.5 & 6, 7 \\
                   & 1998 Jul 26 & 6.45082(1) & 16.7 $\pm$ 0.2 & \ &7\\
 AXP 1841-05...... & 1993 Oct 11 & 11.76668(6)&41.3$\pm$0.1 & 3.4 & 8,9\\
                   &1998, Mar 27 & 11.77243(7)&41.3$\pm$0.1& \ &9\\
 AXP 2259+59...... & 1993 May 30&6.97884(2)&0.4883$\pm$0.0003& 3.9 &10,11 \\
                   &1995, Aug 11 &
                   6.9788793(8)&0.4883$\pm$0.0003& \  &10\\
 AXP 0142+62...... & 1994 Sep 18 & 8.68794(7) & 2.2$\pm 0.2$ & 4.0 & 12  \\
                   &1998 Aug 21 & 8.68828(4) &2.2$\pm$ 0.2& \ &7\\
 AXP 1709-40...... & 1996 Sep 03 &10.99758(6) & 22 $\pm$ 6  & 2.9 & 13,14\\

\hline
\end{tabular}
\end{center}
$^a$Measured period (1 $\sigma$ error in last
digit).\\
$^b$Assumed period derivative (from references).\\
$^c$BB+PL stands for black-body plus power-law spectral fit \\
$^d$ References.-(1)Hurely et al. 1999; (2)Woods et al. 1999; (3)Murakami et al. 1999;
  (4)Sonobe et al. 1994; (5)Woods et al. 2000; (6)Corbet \& Mihara 1997; (7)Paul et al. 2000
  ; (8)Gotthelf \& Vasisht 1997; (9)Gotthelf, Vasisht, \& Dotani 1999; (10)Kaspi, Chakrabarty, \& Steinberger 1999;
  (11)Cobet et al. 1995; (12)White et al. 1996; (13)Sugizaki et al. 1997; (14)Israel et al. 1999 \\

\end{table*}

%
\clearpage

\clearpage

\clearpage

\end{document}